\pdfoutput=1
\documentclass[prb,twocolumn,showpacs,amsmath,amssymb,floatfix,superscriptaddress]{revtex4-1}

\usepackage{graphicx}%Include figure files
\usepackage{dcolumn}%Align table columns on decimal point
\usepackage{bm}% bold math
\newcommand*{\Tr}{\mathrm{Tr}}
\usepackage{hyperref}

\begin{document}
	
	\title{Antidamping spin-orbit torque driven by spin-flip reflection mechanism on the surface of a topological insulator: A time-dependent nonequilibrium Green function approach}	
	\author{Farzad Mahfouzi}
	\email{farzad.mahfouzi@gmail.com}
	\affiliation{Department of Physics, California State University, Northridge, CA 91330-8268, USA}
	\author{Branislav K. Nikoli\'{c}}
	\affiliation{Department of Physics and Astronomy, University of Delaware, Newark, DE 19716-2570, USA}
	\author{Nicholas Kioussis}
	\affiliation{Department of Physics, California State University, Northridge, CA 91330-8268, USA}
	
	\begin{abstract}
		Motivated by recent experiments observing spin-orbit torque (SOT) acting on the magnetization $\vec{m}$ of a ferromagnetic (F) overlayer on the surface of a three-dimensional topological insulator (TI), we investigate the origin of the SOT and the magnetization dynamics in such systems. We predict that lateral F/TI bilayers  of finite length, sandwiched between two normal metal leads, will generate a large antidamping-like SOT per very low charge current injected parallel to the interface. The large values of antidamping-like SOT  are  {\it spatially localized} around the transverse edges of the F overlayer. Our analysis is based on adiabatic expansion (to first order in $\partial \vec{m}/\partial t$) of time-dependent nonequilibrium Green functions (NEGFs), describing electrons pushed  out of equilibrium both by the applied bias voltage and by the slow variation of a classical degree of freedom [such as $\vec{m}(t)$]. From it we extract formulas for spin torque and charge pumping, which show that they are reciprocal effects to each other, as well as Gilbert damping in the presence of SO coupling. The NEGF-based formula for SOT naturally splits into four components, determined by their behavior (even or odd) under the time and bias voltage reversal. Their complex angular dependence is delineated and employed within Landau-Lifshitz-Gilbert simulations of magnetization dynamics in order to demonstrate capability of the predicted SOT to efficiently switch $\vec{m}$ of a perpendicularly magnetized F overlayer.
	\end{abstract}
	
	\maketitle
	
\section{Introduction}\label{sec:intro}

The spin-orbit torque (SOT) is a recently discovered phenomenon~\cite{Miron2011,Garello2013,Liu2012,Liu2012c} in ferromagnet/heavy-metal (F/HM) lateral heterostructures involves unpolarized charge current injected parallel to the F/HM interface induces  switching or steady-state precession~\cite{Liu2013} of magnetization in the F overlayer. Unlike conventional spin-transfer torque (STT) in spin valves and magnetic tunnel junction (MTJs),~\cite{Ralph2008,Stiles2002,Theodonis2006} where one F layer acts as spin-polarizer of electrons that transfer torque to the second F layer when its free magnetization is noncollinear to the direction of incoming spins, heterostructures exhibiting SOT use a {\it single} F layer. Thus, in F/HM bilayers,  spin-orbit coupling (SOC) at the interface or in the bulk of the HM layer is crucial to spin-polarized injected current via the Edelstein effect (EE)~\cite{Edelstein1990,Aronov1989} or the spin Hall effect (SHE),~\cite{Vignale2010,Sinova2015} respectively.

\begin{figure}
		\includegraphics[scale=0.5,angle=0]{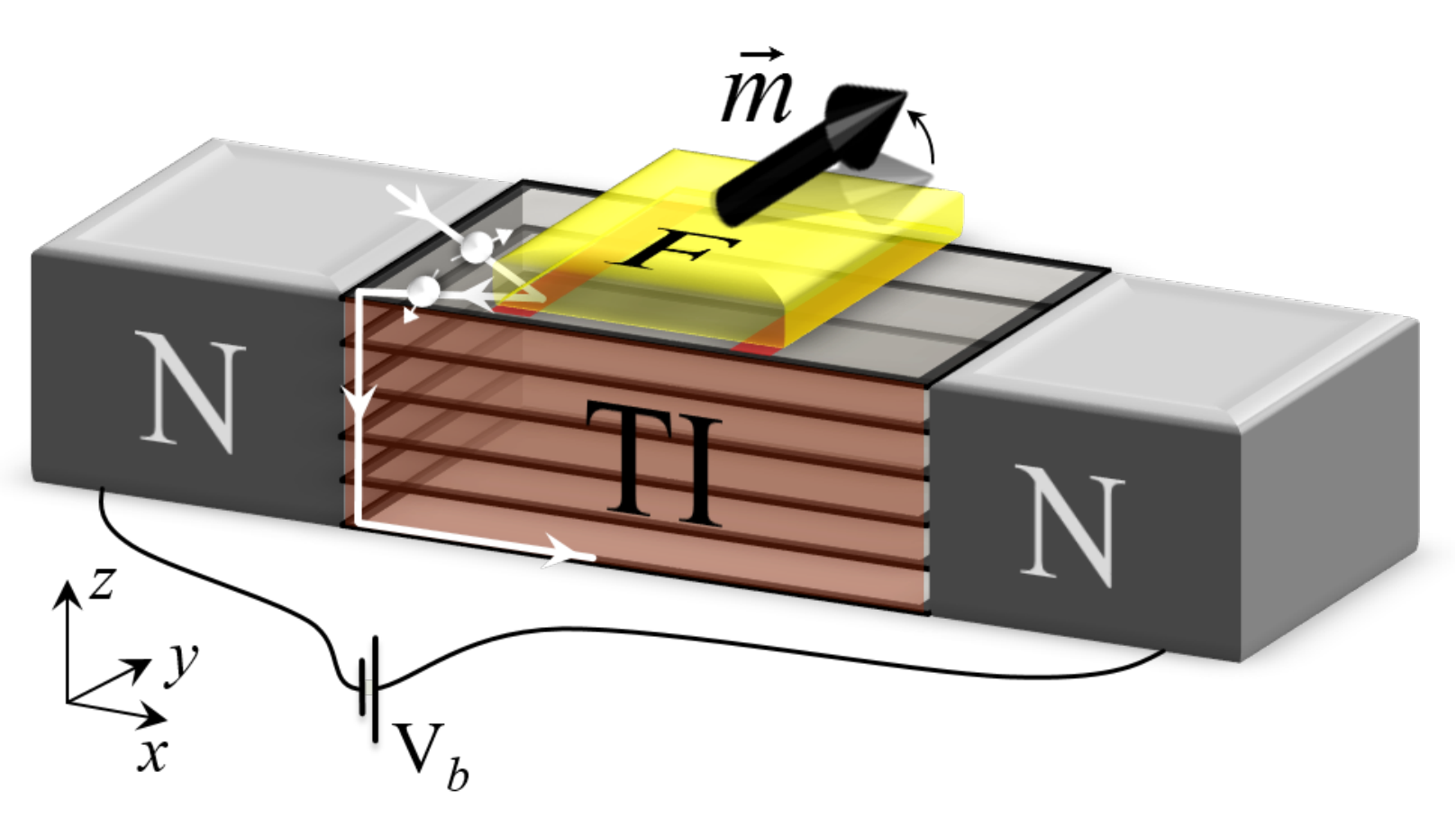}
		\caption{(Color online) Schematic view of F/TI lateral bilayer operated by SOT. The F overlayer has finite length $L_x^F$ and $\vec{m}$ is the unit vector along its free magnetization. The TI layer is attached to two N leads which are semi-infinite in the $x$-direction and terminate into macroscopic reservoirs. We also assume that F and TI layers, as well as N leads, are infinite in the $y$-direction. The unpolarized charge current is injected by the electrochemical potential difference between the left and the right macroscopic reservoirs which sets the bias voltage,  $\mu_L-\mu_R=eV_b$. We mention that the results do not change if the TI surface is covered by the F overlayer partially or fully. }
		\label{fig:fig1}
\end{figure}

The SOT offers potentially more efficient magnetization switching than achieved by using MTJs underlying present STT-magnetic random access memories (STT-MRAM).~\cite{Locatelli2014} Thus, substantial experimental and theoretical efforts have been focused on identifying physical mechanisms behind SOT  whose understanding would pave the way to maximize its value by using optimal materials combinations. For example, very recent experiments~\cite{Mellnik2014,Fan2014a,Wang2015} have replaced HM with three-dimensional topological insulators (3D TIs).~\cite{Hasan2010} The TIs  enhance~\cite{Chang2015,Pesin2012, Garate2010} (by a factor $\hbar v_F/\alpha_R$, where $v_F$ is the Fermi velocity on the surface of  TI and $\alpha_R$ is the Rashba SOC strength~\cite{Park2013,Manchon2015} at the F/HM interface) the transverse nonequilibrium spin density driven by the longitudinal charge current, which is responsible for the large {\it field-like} SOT component~\cite{Garate2010,Yokoyama2011} observed experimentally.~\cite{Mellnik2014,Fan2014a,Wang2015}

Furthermore, recent experiments have also observed antidamping-like SOT in F/TI heterostructures with surprisingly large figure of merit (i.e., antidamping torque per unit applied charge current density) that surpasses~\cite{Mellnik2014,Fan2014a,Wang2015} those measured in a variety of F/HM heterostructures. This component competes against the Gilbert damping which tries to restore magnetization to equilibrium, and its large figure of merit is, therefore, of particular importance for increasing efficiency of magnetization switching. Theoretical understanding of the physical origin of antidamping-like SOT is crucial to resolve the {\it key challenge} for anticipated applications of SOT generated by TIs---demonstration of  magnetization switching of the F overlayer at room temperature (thus far, magnetization switching has been demonstrated only at cryogenic temperature~\cite{Fan2014a}). 
 
However, the microscopic mechanism behind its large  magnitude~\cite{Mellnik2014,Fan2014a,Wang2015} and ability to efficiently (i.e., using as little dc current density as possible) switch magnetization~\cite{Fan2014a} remains under scrutiny. For example, TI samples used in these experiments are often unintentionally doped, so that bulk  charge carriers can generate  antidamping-like SOT via rather large~\cite{Jamali2015} SHE (but not sufficient to explain all reported values~\cite{Mellnik2014,Fan2014a}). The simplistic picture,~\cite{Mellnik2014} in which electrons spin-polarized by the EE  diffuse into the F overlayer~\cite{Mellnik2014} to deposit spin angular momentum within it, cannot operate in technologically relevant F overlayers of \mbox{$\simeq 1$} nm thickness~\cite{Wang2015} or explain complex angular dependence~\cite{Fan2014a,Garello2013,Lee2015} typically observed for  SOT. The Berry curvature mechanism~\cite{Lee2015,Li2015} for antidamping-like SOT applied to lateral F/TI heterostructures predicts its peculiar dependence on the magnetization orientation,~\cite{Ndiaye2015} vanishing when magnetization $\vec{m}$ is parallel to the F/TI interface. This feature has thus far not been observed experimentally,~\cite{Fan2014a} and, furthermore, it makes such antidamping-like SOT less efficient~\cite{Ndiaye2015} (by requiring larger injected currents to initiate magnetization switching) than standard SHE-driven~\cite{Liu2012,Liu2012c} antidamping-like SOT.

We note that the recent experimental~\cite{Mellnik2014,Fan2014a,Wang2015} and theoretical~\cite{Mellnik2014,Ndiaye2015} studies of SOT in lateral F/TI bilayer have focused on the geometry where an  infinite F overlayer covers an infinite TI layer. Moreover, they assume~\cite{Mellnik2014,Ndiaye2015,Yokoyama2009} purely two-dimensional transport where only the top surface of the TI layer is explicitly taken into account by the low-energy effective (Dirac) Hamiltonian supplemented by the Zeeman term due to the magnetic proximity effect. On the other hand, transport in realistic TI-based heterostructures is always three-dimensional, with {\it unpolarized} electrons being injected from normal metal contacts, reflected from the F/TI edge to flow along the surface of the TI in the $yz$-plane  and then along the bottom TI surface in Fig.~\ref{fig:fig1}. In fact, electrons also flow within a thin layer (of thickness $\lesssim 2$ nm in  Bi$_2$Se$_3$ as the prototypical TI material) underneath the top and bottom surfaces due to top and bottom metallic surfaces of the TI doping the bulk via evanescent wave functions.~\cite{Chang2015} Therefore, in this study we consider more realistic and experimentally relevant~\cite{Sanchez2013} F/TI bilayer geometries, illustrated in Fig.~\ref{fig:fig1}, where the TI layer of finite length $L^\mathrm{TI}_x$ and finite thickness $L^\mathrm{TI}_z$ is (partially or fully) covered by the F overlayer of length $L^\mathrm{F}_x$. The two semi-infinite ideal N leads are directly attached to the TI layer. we should mention that the result does not depend on the length of TI layer that is covered by the FM. 

\begin{figure*}
					\includegraphics[scale=0.6,angle=0]{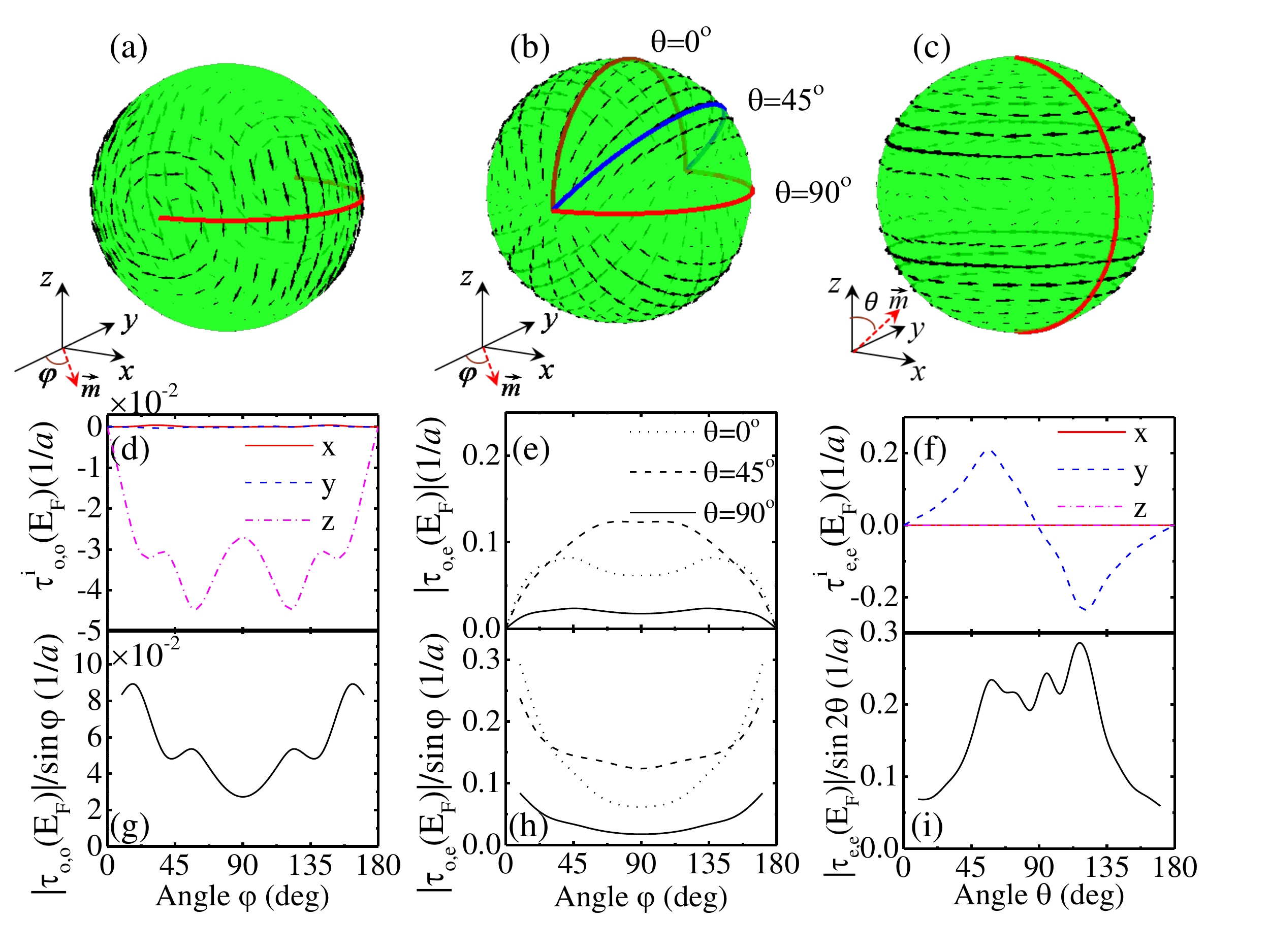}
					\caption{(Color online) (a)--(c) The vector field of SOT components $\vec{\tau}_{\mu,\nu}(E_F)$, defined by Eq.~\eqref{eq:components}, for different directions of $\vec{m}$ on the unit sphere. The angular behavior in (a) and
					(c) shows that  $\vec{\tau}_\mathrm{o,o}$ and $\vec{\tau}_\mathrm{e,e}$ behave as field-like torques, while that 
					in (b) shows that $\vec{\tau}_\mathrm{o,e}$ behaves as antidamping-like torque. Cartesian components  [(d)-(f)] and magnitude of $\vec{\tau}_{\mu,\nu}$ [(g)--(i)]  along the trajectories denoted by solid lines in the corresponding panels (a)-(c). The magnitude of $\vec{\tau}_{\mu,\nu}$ is divided by: $\lvert\vec{m}\times \hat{y}\lvert$ in (g); $\lvert\vec{m}\times(\vec{m}\times\hat{y})\lvert$ in (h) and  $2(\vec{m}\cdot \hat{z})\lvert\vec{m}\times \hat{z}\lvert$ in (i).}
					\label{fig:fig2}
\end{figure*}

Our principal results are twofold and are summarized as follows:  

({\it i}) {\it Theoretical prediction for SOT:} We predict that the geometry in Fig.~\ref{fig:fig1} will generate large antidamping-like SOT per low injected charge current.  By studying spatial dependence of the SOT (see Fig.~\ref{fig:fig4}), we show that in a clean FM/TI interface the electrons exert anti-damping torque on the FM as they enter into the interface and unless interfacial roughness or impurities are included the torque remains mainly concentrated around the edge of the interface. Although the exact results show strong nonperturbative features, based on second order perturbation we present two different interpretations showing that the origin of the antidamping SOT relies on the spin-flip reflection of the chiral electrons injected into the FM/TI interface. Its strong angular dependence (see Fig.~\ref{fig:fig2}), i.e., dependence on the magnetization direction $\vec{m}$,  offers a unique signature that can be used to distinguish it from other possible physical mechanisms. By numerically solving the Landau-Lifshitz-Gilbert (LLG) equation in the macrospin approximation, we demonstrate (see Figs.~\ref{fig:fig5} and ~\ref{fig:fig6}) that the obtained SOT is capable of switching of a single domain magnetization of a perpendicularly magnetized F overlayer with bias voltage in the oder of the Magneto-Crystaline Anisotropy (MCA) energy. 

%The antidamping-like SOT driven by a mechanism unique to the surface of TIs has been overlooked by prior quantum (such as the Kubo %formula~\cite{Haney2013a,Freimuth2014,Li2015}) and semiclassical (such as the Boltzmann equation~\cite{Haney2013}) theories of SOT because they have been %developed for translationally invariant systems. Therefore, they cannot take into account spin-flip reflection processes on the surface of TI coupled to finite-size F overlayer in the geometry of  Fig.~\ref{fig:fig1}(a).

({\it ii}) {\it Theoretical formalism for SOT:}  The widely used quantum (such as the Kubo formula~\cite{Ndiaye2015,Freimuth2014,Lee2015,Li2015}) and semiclassical (such as the Boltzmann equation~\cite{Haney2013}) transport approaches to SOT are tailored for geometries where an infinite F layer covers  an infinite TI or HM layer. Due to translational invariance, the nonequilibrium spin density $\vec{S}$ induced by the EE on the surface of TI or HM layer has uniform orientation $\vec{S}=(0,S_y,0)$ [in the coordinate system in Fig.~\ref{fig:fig1}], which then provides reference direction for defining field-like, 
$\tau_\mathrm{f} \vec{m} \times \hat{y}$, and antidamping-like, $\tau_\mathrm{ad} \vec{m} \times (\vec{m} \times \hat{y})$, components of SOT. In order to analyze spatial dependence of SOT in the device geometry of Fig.~\ref{fig:fig1}, while not assuming anything {\it a priori} about the orientation of  field-like and antidamping-like components of SOT, we employ adiabatic expansion~\cite{Bode2012} of time-dependent nonequilibrium Green functions (NEGFs)~\cite{Stefanucci2013,Haug2008} to derive formulas for torque,  charge pumping~\cite{Mahfouzi2012,Mahfouzi2014a} and Gilbert damping~\cite{Yokoyama2010a} in the presence of SOC. The NEGF-based formula for SOT naturally splits into four components, determined by their behavior (even or odd) under the time and bias voltage reversal. This gives us a general framework in quantum mechanics to analyze the dissipative (antidamping-like) and nondissipative (field-like) force (torque) vector fields for a set of canonical variables (magnetization directions). Their angular (see Fig.~\ref{fig:fig2}) and spatial (see Fig.~\ref{fig:fig4}) dependence shows that although field-like and antidamping-like  SOTs are predominantly along the $\vec{m} \times \hat{y}$ and $\vec{m} \times (\vec{m} \times \hat{y})$ directions, respectively, they are not uniform and can exhibit significant deviation from the trivial angular dependence defined by these cross products [see Fig.~\ref{fig:fig2}(h)]. 

%The importance of our formalism for SOT is also highlighted by the fact that previous studies~\cite{Yokoyama2011} of SOT in a similar step-like setup, based on simplistic scattering theory, were not able to identify a non-zero antidamping-like component of SOT. 

The paper is organized as follows. In Sec.~\ref{sec:negf}, we present the adiabatic expansion of time-dependent NEGFs, in a representation that is alternative to  Wigner representation~\cite{Haug2008} (usually employed for this type of derivation~\cite{Bode2012}), and derive expressions for torque, charge pumping and Gilbert damping. In Sec.~\ref{sec:angular}, we decompose the NEGF-based expression for SOT into four components, determined by their behavior (even or odd) under the time and bias voltage reversal, and investigate their angular dependence. Section~\ref{sec:transmission} discusses the angular dependence of the zero-bias  transmission function which identifies the magnetization directions at which substantial reflection occurs. In Sec.~\ref{sec:spatial}, we study spatial dependence of SOT components and discuss their physical origin.  Section~\ref{sec:llg} presents LLG simulations of  magnetization dynamics in the presence of predicted SOT, as well as a switching phase  diagram of the magnetization state as a function of the in-plane external magnetic field and SOT. We conclude in Sec.~\ref{sec:conclusion}.

\section{Theoretical Formalism}\label{sec:negf}

We first describe the time-dependent Hamiltonian model, $\mathbf{H}(t) = \mathbf{H}_0  + \mathbf{U}(t)$, of the lateral F/TI heterostructure in Fig.~\ref{fig:fig1}. Here $\mathbf{H}_0$ is the minimal tight-binding model for 3D TIs like Bi$_2$Se$_3$ on a cubic lattice of spacing $a$ with four orbitals per site.~\cite{Liu2010} The thickness,  $L_z^\mathrm{TI}=8a$ of the TI layer is sufficient to prevent hybridization between its top and bottom metallic surface states.~\cite{Chang2015} The time-dependent potential
\begin{equation}\label{eq:proximity}
\mathbf{U}(t) = - {\Delta}_\mathrm{surf} \mathbf{1}_m \vec{m}(t) \cdot \vec{\bm \sigma}/2,
\end{equation}
depends on time through the magnetization of the F overlayer which acts as the slowly varying classical degree of freedom. Here $\vec{m}(t)$ is the unit vector along the direction of magnetization, \mbox{$\Delta_\mathrm{surf}=0.28$ eV} is the proximity induced exchange-field term and $\mathbf{1}_m$ is a diagonal matrix with elements equal to unity for sites within the F/TI contact region in Fig.~\ref{fig:fig1} and zero elsewhere.  The semi-infinite ideal N leads in Fig.~\ref{fig:fig1} are taken into account through the self-energies~\cite{Stefanucci2013,Haug2008} ${\bm \Sigma}_{L,R}$ computed for a tight-binding model with one spin-degenerate orbital per site. The details of how to properly couple ${\bm \Sigma}_{L,R}$ to $\mathbf{H}_0$, while taking into account that the spin operators for electrons on the Bi and Se sublattices of the TI are inequivalent,~\cite{Silvestrov2012} can be found in Ref.~\onlinecite{Chang2014a}.

Within the NEGF formalism~\cite{Stefanucci2013,Haug2008} the advanced and lesser GFs matrix elements of the tight-binding Hamiltonian, $\mathbf{H}_0$, are defined by $G_{{\bf ii'},{oo'},{ss'}}(t,t') = - i \Theta(t-t') \langle \{\hat{c}_{{\bf i}o s}(t) , \hat{c}^\dagger_{{\bf i'} o' s'}(t')\}\rangle$, and $G^<_{{\bf ii'},{oo'},{ss'}}(t,t') = i \langle \hat{c}^\dagger_{{\bf i'}o's'}(t') \hat{c}_{{\bf i}o s}(t) \rangle$, respectively. Here, $\hat{c}_{{\bf i}o s}^\dagger$ ($\hat{c}_{{\bf i}o s}$) is the creation (annihilation) operator for an electron on site, $\mathbf{i}$, with orbital, $o$, and spin $s$, respectively, $\langle \ldots \rangle$ denotes the nonequilibrium statistical average, and $\hbar=1$ to simplify the notation. These GFs are the matrix elements of the corresponding matrices $\mathbf{G}$ and $\mathbf{G}^<$ used throughout the text.

Under stationary conditions, the two GFs depend on the difference of the time arguments, $t-t'$, and can be Fourier transformed to energy. In the strictly adiabatic  limit one can employ~\cite{Salahuddin2006} the same retarded GF, \mbox{$\mathbf{G}_t(E) = [E-\mathbf{H}(t) - {\bm \Sigma}_L - {\bm \Sigma}_R]^{-1}$}, as under stationary conditions, but where the GF depends parametrically on time (denoted by the subscript $t$) and is computed for the frozen-in-time configuration of $\mathbf{U}(t)$. However, even for slow evolution of $\vec{m}(t)$  corrections~\cite{Bode2012} to the adiabatic GF are needed to describe dissipation effects such as Gilbert damping or the charge current which can be pumped~\cite{Mahfouzi2012} by the dynamics of $\vec{m}(t)$.

The so-called adiabatic expansion, which yields corrections beyond the strictly adiabatic limit, is traditionally performed using the Wigner representation~\cite{Haug2008} in which the fast and slow time scales are easily identifiable.~\cite{Bode2012} The slow motion implies that the NEGFs vary slowly with the central time $t_c=(t+t')/2$ while they change fast with
the relative time $t_r=t-t'$. By expanding the Wigner transformation of NEGFs
\begin{equation}
\mathbf{G}_W^{(<)}(E,t_c)=\int_{-\infty}^{\infty} dt_r \, e^{i E t_r}  \mathbf{G}^{(<)} \left(t_c+\frac{t_r}{2},t_c-\frac{t_r}{2} \right),
\end{equation}
in the central time $t_c$ while keeping only terms containing first-order derivatives $\partial/\partial t_c$ (due to the slow variation with $t_c$) gives the first-order correction beyond the strictly adiabatic limit.~\cite{Bode2012} This route requires to handle complicated expressions resulting from the  Wigner transform applied to convolutions of the type $C(t_1,t_2) = \int dt_3\, C_1(t_1,t_3)C_2(t_3,t_2)$.

Here we provide an alternative derivation of the first-order nonadiabatic correction. Namely, we consider $t$ (observation time) and $t-t'$ (relative time) as the natural variables to describe the time evolution of NEGFs and then perform the following Fourier transform~\cite{Arrachea2007}
\begin{equation}\label{eq:ft}
\mathbf{G}(t,t')=\int_{-\infty}^\infty \frac{dE}{2\pi} e^{iE(t-t')} \mathbf{G}(E,t).
\end{equation}
The standard equations of motion for $\mathbf{G}(t,t')$ and $\mathbf{G}^<(t,t')$ are cumbersome to manipulate~\cite{Arrachea2007,Jauho1994} or solve numerically,~\cite{Gaury2014} so they are usually transformed to some other representation.\cite{Mahfouzi2012}  Here we replace $\mathbf{G}(t,t')$ in the standard equations of motion with the rhs of Eq.~\eqref{eq:ft} to arrive at:
\begin{align}
&\left[ \left(E-i\frac{\partial}{\partial t}\right) \bold{1}-\bold{H}_0-\mathbf{U}(t)-\boldsymbol{\Sigma}\left(E-i\frac{\partial}{\partial t} \right) \right]{\bold{G}}(E,t)={\bold{1}},
\end{align}
and
\begin{align}
&\mathbf{G}^{<}(t,t')= \int \frac{dE}{2\pi} \, \bold{G}(E,t) {\bm \Sigma}^<(E) \bold{G}^{\dagger}(E,t')e^{iE(t-t')}.
\end{align}
For the two-terminal heterostructure in Fig.~\ref{fig:fig1} in the elastic transport regime,~\cite{Stefanucci2013,Haug2008} ${\bm \Sigma}(E)={\bm \Sigma}_L(E)+{\bm \Sigma}_R(E)$ and ${\bm \Sigma}^<(E)=if_L(E){\bm \Gamma}_L(E) + if_R(E){\bm \Gamma}_R(E)$, where $\mathbf{\Gamma}_{L,R} = i (\mathbf{\Sigma}_{L,R}-\mathbf{\Sigma}_{L,R}^\dagger)$. The Fermi-Dirac distribution functions of electrons in the macroscopic reservoirs into which the left and right N leads terminate are $f_{L,R}(E)=f(E-\mu_{L,R})$, where the difference between the electrochemical potentials, $\mu_{L,R}=E_F+eV_{L,R}$, defines the bias voltage $eV_b = \mu_L-\mu_R$.

\begin{widetext}
Using the following identity
\begin{equation}
\sum_{\alpha=L,R} i\bold{G}\boldsymbol{\Gamma}_\alpha \bold{G}^{\dagger} =(\bold{G}-\bold{G}^{\dagger})+i\frac{\partial}{\partial E}\left(\bold{G}\frac{\partial\boldsymbol{U}}{\partial t}\bold{G}^{\dagger}\right)+\mathcal{O} \left( \frac{\partial^2 \mathbf{U}}{\partial t^2} \right),
\end{equation}
the lesser GF to first order in small $\partial \mathbf{U}(t)/\partial t$ and in the low bias $V_b$ regime (i.e., the linear-response transport regime) can be expressed as,
\begin{equation}\label{eq:glesser}
\mathbf{G}^{<}(t,t) \simeq \int \frac{dE}{2\pi} \left([\mathbf{G}(E,t) - \mathbf{G}^\dagger(E,t)] f(E)  + \sum_{\alpha=L,R}  f' eV_\alpha \mathbf{G}_t \mathbf{\Gamma}_\alpha \mathbf{G}_t^\dagger  + i f' \mathbf{G}_t \frac{\partial \mathbf{U}(t)}{\partial t} \mathbf{G}_t^\dagger\right),
\end{equation}
where, the first term corresponds to the density matrix of the equilibrium electrons occupying the time dependent single particle states, while the second and third terms describe the density matrix of the excited (nonequilibrium) electrons occupying the states close to the Fermi energy due to the bias voltage and time dependent term in the Hamiltonian, respectively. The retarded GF in the first term in Eq.~\eqref{eq:glesser} expanded to first order in $\partial \mathbf{U}(t)/\partial t$ is of the form
\begin{equation}\label{eq:retarded}
\mathbf{G}(E,t) \simeq \mathbf{G}_t + i \frac{\partial \mathbf{G}_t}{\partial E} \frac{ \partial \mathbf{U}(t)}{\partial t} \mathbf{G}_t.
\end{equation}
 The lesser GF determines the time-dependent nonequilibrium density matrix
\begin{equation}
{\bm \rho}(t) = \frac{1}{i} \mathbf{G}^<(t,t),
\end{equation}
from which we determine the time-dependent expectation values of physical observables, $A(t) = \mathrm{Tr}\, [{\bm \rho}(t) \hat{A}]$. In particular, the relevant quantities for the heterostructure in Fig.~\ref{fig:fig1} are the charge current			
	\begin{eqnarray} \label{eq:current}	
	I_{\alpha}(t)  & = & e \int \frac{dE}{2\pi i}\, \Tr \left\{ \boldsymbol{\Gamma}_{\alpha}(E) \bold{G}^{<}(t,t) + f_{\alpha}(E) \boldsymbol{\Gamma}_{\alpha}(E) [ \bold{G}(E,t)-\bold{G}^\dagger(E,t)] \right\} \nonumber \\
	& = & \frac{e^2}{2 \pi} \int dE\, f'(E) \left\{ \sum_{\beta} (V_{\beta}-V_{\alpha}) T^{\alpha\beta}(E) -  \frac{\Delta_\mathrm{surf}}{2e} \sum_i \frac{\partial m_i}{\partial t}T^{\alpha i}(E) \right\},
	\end{eqnarray}
	and the spin density
	\begin{equation}  \label{eq:spin}
	s^i(t)  =  \int \frac{dE}{2\pi i}\Tr\left[{\bm \sigma}_i \mathbf{1}_m \bold{G}^{<}\right] = \int  \frac{dE}{2\pi} \left\{ \sum_{\alpha} f_{\alpha}(E) T^{i\alpha}(E) -\frac{\Delta_\mathrm{surf}}{2} \sum_j f'(E)\frac{\partial m_j}{\partial t}T^{ij}(E) \right\},
	\end{equation}
where $\alpha,\beta \in \{L,R\}$ and $i,j \in \{x,y,z\}$.
\end{widetext}

The ``trace-formulas'' in Eqs.~\eqref{eq:current} and ~\eqref{eq:spin} 
	\begin{subequations}\label{eq:trans}
	\begin{eqnarray}\label{eq:transcharge}
	&&T^{\alpha\beta}(E) =  \Tr\left[\boldsymbol{\Gamma}_{\alpha} \bold{G}_t \boldsymbol{\Gamma}_{\beta} \bold{G}_t^\dagger \right], \\
	&&T^{\alpha i}(E) = \Tr\left[ {\bm 1}_m {\bm \sigma}_i \bold{G}_t^\dagger {\bm \Gamma}_{\alpha} \bold{G}_t \right], \\
	&&T^{i\alpha }(E)  = \Tr\left[ {\bm 1}_m {\bm \sigma}_i \bold{G}_t {\bm \Gamma}_{\alpha} \bold{G}_t^\dagger\right], \\
	&&T^{ij}(E) = \Tr \left[{\bm 1}_m{\bm \sigma}_i (\mathbf{G}_t^\dagger - \mathbf{G}_t) {\bm 1}_m{\bm \sigma}_j (\mathbf{G}_t - \mathbf{G}_t^\dagger) \right],
	\end{eqnarray}
	\end{subequations}
determine charge current~\cite{Caroli1971} due to $V_b$, charge current pumped~\cite{Mahfouzi2012} by the dynamics of $\vec{m}(t)$ in the presence of SOC, the spin torque, and Gilbert damping tensor, respectively. In the expression for the spin density we ignore the antisymmetric part of $T^{ij}$ which corresponds to the renormalization of the precession frequency of magnetization dynamics.~\cite{Bode2012} Note that $T^{\alpha i}$  in Eq.~\eqref{eq:trans}(b) and its time-reversal $T^{i \alpha}$ in Eq.~\eqref{eq:trans}(c) reveal a {\it reciprocal}~\cite{Freimuth2015} relation between charge pumping by magnetization dynamics in the presence of SOC and current-driven SOT at {\em each} instant of time $t$.

The spin density in Eq.~\eqref{eq:spin} enters into the LL equation for the magnetization dynamics
		\begin{align}\label{eq:LLG1}
		\frac{\partial \vec{m}}{\partial t}=-\frac{{\Delta}_\mathrm{surf}}{2} \vec{m}\times \vec{s}(t).
		\end{align}
		The first term in Eq.~\eqref{eq:spin} generates the spin torque in Eq.~\eqref{eq:LLG1} which has three contributions: ({\em i}) an equilibrium component responsible for the interlayer exchange interaction in the presence of a second F layer and/or magneto-crystalline anisotropy (MCA) in the presence of SOC; ({\em ii}) a bias-induced field-like torque modifying the equilibrium interlayer  exchange and MCA fields; and ({\em iii}) a damping (antidamping)-like torque describing angular momentum loss (gain) due to the flux of Fermi surface electrons.

\section{Angular dependence of SOT components}\label{sec:angular}

In order to understand the different contributions to  $T^{i\alpha}(E)$, we decompose it into even (e) or odd (o) terms under time-reversal  $\bold{G}_t \mapsto \bold{G}^{\dagger}_t$:
\begin{eqnarray}
T_\mathrm{e}^{i\alpha}(E) & = & [T^{i\alpha}(E)+T^{\alpha i}(E)]/2, \\
T_\mathrm{o}^{i\alpha}(E) & = & [T^{i\alpha}(E)-T^{\alpha i}(E)]/2. 
\end{eqnarray}
Since $\sum_{\alpha}T_\mathrm{o}^{i\alpha}(E)=0$, the contribution of the odd  component to the equilibrium spin density in Eq.~\eqref{eq:spin} vanishes identically, while its nonzero values appear only for $E$ within the bias window around $E_F$. This motivates further splitting of $T^{i\alpha}(E)$  into four components for the case of two-terminal devices
	\begin{subequations}\label{eq:components}
		\begin{eqnarray}
		T_\mathrm{e,\nu}^{i}(E) & = & \frac{T_\mathrm{\nu}^{iL}(E)+T_\mathrm{\nu}^{iR}(E)}{2},\\
		%	T_\mathrm{\mu,o}^{i}(E) & = & \frac{T_\mathrm{o}^{iL}(E)+T_\mathrm{o}^{iR}(E)}{2} \equiv 0, \\
		T_\mathrm{o,\nu}^{i}(E) & = & \frac{T_\mathrm{\nu}^{iL}(E)-T_\mathrm{\nu}^{iR}(E)}{2}, 
		%	T_\mathrm{o,o}^{i}(E) & = & \frac{T_\mathrm{o}^{iL}(E)-T_\mathrm{o}^{iR}(E)}{2},
		\end{eqnarray}
where the first and second subscripts denote their behavior (even or odd) under bias reversal $V_b \mapsto - V_b$  and time reversal, respectively. The corresponding four components of torque are determined by
    \begin{eqnarray}
    \vec{\mathcal{T}}_{\mathrm{e},\nu} & = & \int \frac{dE}{2\pi} \,  [f_L(E) + f_R(E)] \vec{\tau}_{\mathrm{e},\nu}(E), \label{eq:mca} \\
    \vec{\mathcal{T}}_{\mathrm{o},\nu} & = & \int \frac{dE}{2\pi} \,  [f_L(E) - f_R(E)] \vec{\tau}_{\mathrm{o},\nu}(E), \label{eq:stt}
    \end{eqnarray}
	\end{subequations}
where the energy-resolved torque is given by 
\begin{equation}
\vec{\tau}_{\mu,\nu}(E)=-\frac{\Delta_\mathrm{surf}}{2} \vec{m} \times \vec{T}_{\mu,\nu}(E),
\end{equation}
and $\mu,\nu \in \{ \mathrm{e},\mathrm{o} \}$. The terms $\vec{\mathcal{T}}_{\mathrm{o},\mathrm{o}}$ and  $\vec{\mathcal{T}}_{\mathrm{o},\mathrm{e}}$ are non-zero only in nonequilibrium driven by $V_b \neq 0$, and depend on electronic states in the bias voltage window around the Fermi energy (or on the Fermi surface states in the linear-response regime where integrals are avoided by multiplying integrand by $eV_b$). The term $\vec{\mathcal{T}}_{\mathrm{e},\mathrm{o}} \equiv 0$ is zero, while $\vec{\mathcal{T}}_{\mathrm{e},\mathrm{e}}$ is nonzero also in equilibrium and, therefore, depends on all occupied electronic states.

Figures~\ref{fig:fig2}(a-c) show the {\it net} vector field (summed over all sites of the F overlayer) of $\vec{\tau}_{\mu,\nu}(E_F)$ at zero temperature for different directions of $\vec{m}$ on the unit sphere. Their angular behavior reveals that: (\textit{i}) $\vec{\tau}_\mathrm{o,o}$ shown in Fig.~\ref{fig:fig4}(a) is the field-like SOT generated by the EE, with predominant orientation  along the $\vec{m}\times\hat{y}$-direction; (\textit{ii})  $\vec{\tau}_\mathrm{o,e}$ in Fig.~\ref{fig:fig4}(b) is the  antidamping-like SOT with predominant orientation along the $\vec{m}\times(\vec{m}\times \hat{y})$-direction; and (\textit{iii})  $\vec{\tau}_\mathrm{e,e}$ in Fig.~\ref{fig:fig4}(c) along the $\vec{m}\times\hat{z}$ direction is the field-like component whose angular dependence behaves approximately as \mbox{$2(\vec{m}\cdot  \hat{z})|\vec{m}\times\hat{z}|\equiv\sin(2\theta)$} typical for torque components generated by the MCA field.~\cite{Garello2013}

The corresponding angular dependence of the {\it net} $\tau^i_\mathrm{o,o} (i\in\{x,y,z\})$, $|\vec{\tau}_\mathrm{o,e}|$, and $\tau^i_\mathrm{e,e}$ along the solid trajectories shown in Figs. ~\ref{fig:fig2}(a-c) are plotted in Figs. ~\ref{fig:fig2}(d-f), respectively. We find that the magnitude of the maximum antidamping-like SOT is about a factor of four larger than that of the field-like SOT. Additionally, the field-like SOT peaks when the magnetization is in-plane.  In contrast, the antidamping-like SOT peaks when the magnetization is out of plane, which can be attributed to the gap opening of the Dirac cone which in turn enhances the reflection (see Sec.~\ref{sec:transmission}) at the lateral boundaries of the F overlayer.

\begin{figure}
			\includegraphics[scale=0.7,angle=0]{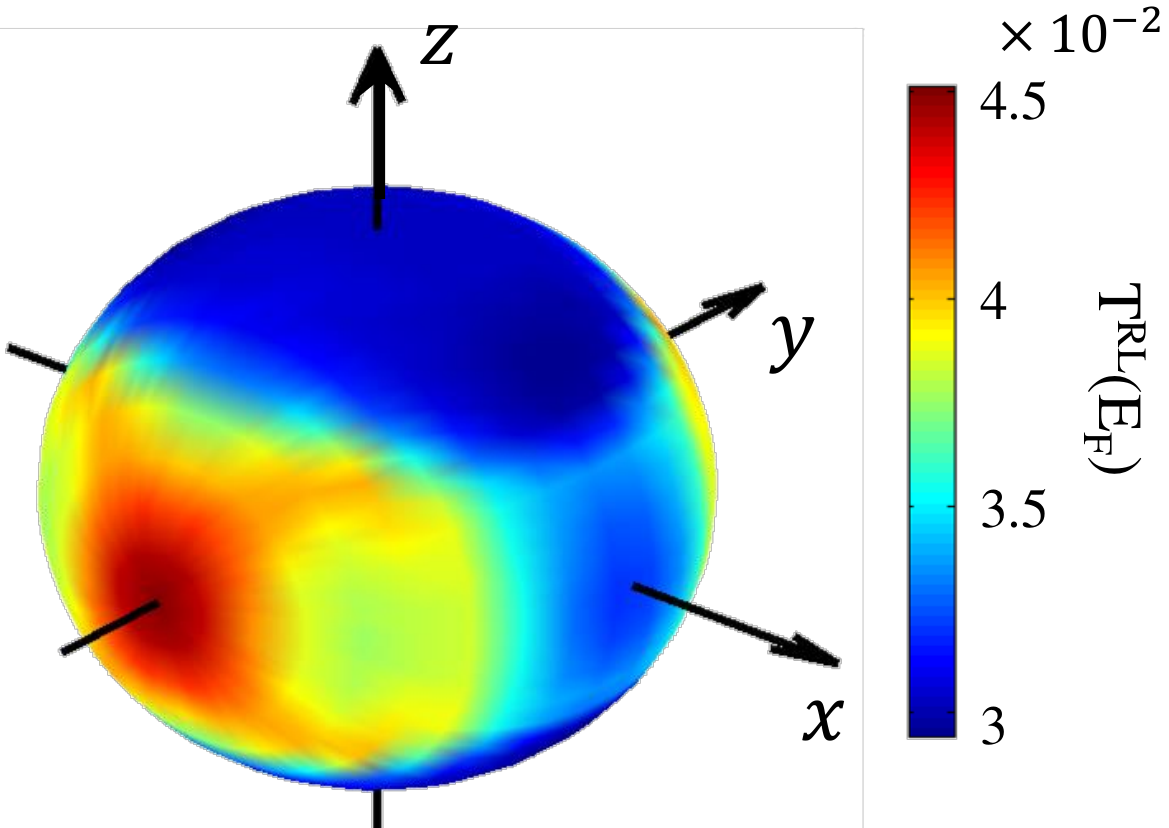}
			\caption{(Color online) The transmission function $T^{RL}(E_F)$ in Eq.~\eqref{eq:transcharge} for two-terminal heterostructure shown in Fig.~\ref{fig:fig1} at different directions of magnetization $\vec{m}$ on the unit sphere. The Fermi energy is set at \mbox{$E_F=3.1$ eV} (which is $0.1$ eV above the DP).}
			\label{fig:fig3}
\end{figure}

Note that the magnitude of the {\it net} SOT components shown in Figs. ~\ref{fig:fig2}(g-i) exhibits strong
angular dependence because of the large SOC on the TI surface similar to that found in F/HM heterostructures when the Rashba SOC at the interface is sufficiently strong.~\cite{Lee2015,Li2015} In particular, the significant deviation of the angular dependence of the antidamping-like SOT from the trivial
$|\vec{m}\times(\vec{m}\times \hat{y})|$ behavior in the limit of $\vec{m}\rightarrow\hat{y}$, shown quantitatively in Fig. ~\ref{fig:fig2}(h), indicates its nonperturbative variation with respect to the magnetization direction. A similar nonperturbative angular behavior (i.e., strong deviation from the standard $\propto \sin^2 \theta$ dependence on the precession cone angle $\theta$) has been found for adiabatic charge pumping from a precessing F overlayer attached to the edge of 2D TIs~\cite{Qi2008,Mahfouzi2010} or to the surface of 3D TIs.~\cite{Mahfouzi2014a}

\section{Angular dependence of transmission function}\label{sec:transmission}
Figure~\ref{fig:fig3} shows the transmission function $T^{RL}(E_F)$ in Eq.~\eqref{eq:transcharge} for the heterostructure in Fig.~\ref{fig:fig1} versus the orientation of $\vec{m}$ on the unit sphere. We find that the charge current determined by $T^{RL}(E)$  is smallest~\cite{Kong2011} when $\vec{m} \parallel \hat{z}$ or $\vec{m} \parallel \hat{x}$. This is due to the reflection of Dirac electrons on the TI surface from the lateral boundaries of the F overlayer. Underneath the F overlayer, the exchange field, $-{\Delta}_\mathrm{surf} \vec{m} \cdot \vec{\bm \sigma}/2$, induced by the magnetic proximity effect is superimposed on the Dirac cone surface dispersion. This opens an energy gap ${\Delta}_\mathrm{surf}$ when $\vec{m} \parallel \hat{z}$ (or smaller gap ${\Delta}_\mathrm{surf} \cos \theta$ for $m_z \neq 0$) at the DP of the TI region underneath the F overlayer, which in turn gives rise to strong electronic reflection when $\vec{m} \parallel \hat{z}$. For $\vec{m} \parallel \hat{x}$, there is no energy gap at the DP and the Dirac cone effectively shifts away from the center of the Brillouin zone due to proximity exchange field. Nevertheless electrons polarized by the EE along the $y$-axis reflect from the  magnetization pointing along the $x$-axis.

%Note that the current always remains finite, even when $E_F$ lies very close the DP (where the density of states of bulk TI vanishes), because of evanescent wavefunctions injected~\cite{Chang2014a} by the metallic N leads into the TI region for heterostructure in Fig.~\ref{fig:fig1} (or other mechanisms responsible for always measured~\cite{Kim2012} minimum conductivity of Dirac materials). In addition, uncovered TI region in Fig.~\ref{fig:fig1} will inject evanescent wavefunctions~\cite{Yokoyama2011,Mahfouzi2010} into the gapped state (when $m_z \neq 0$) within the F/TI contact region.

\section{Spatial dependence of SOT components and physical origin of antidamping-like SOT}\label{sec:spatial}

\begin{figure}
			\includegraphics[scale=0.3,angle=0]{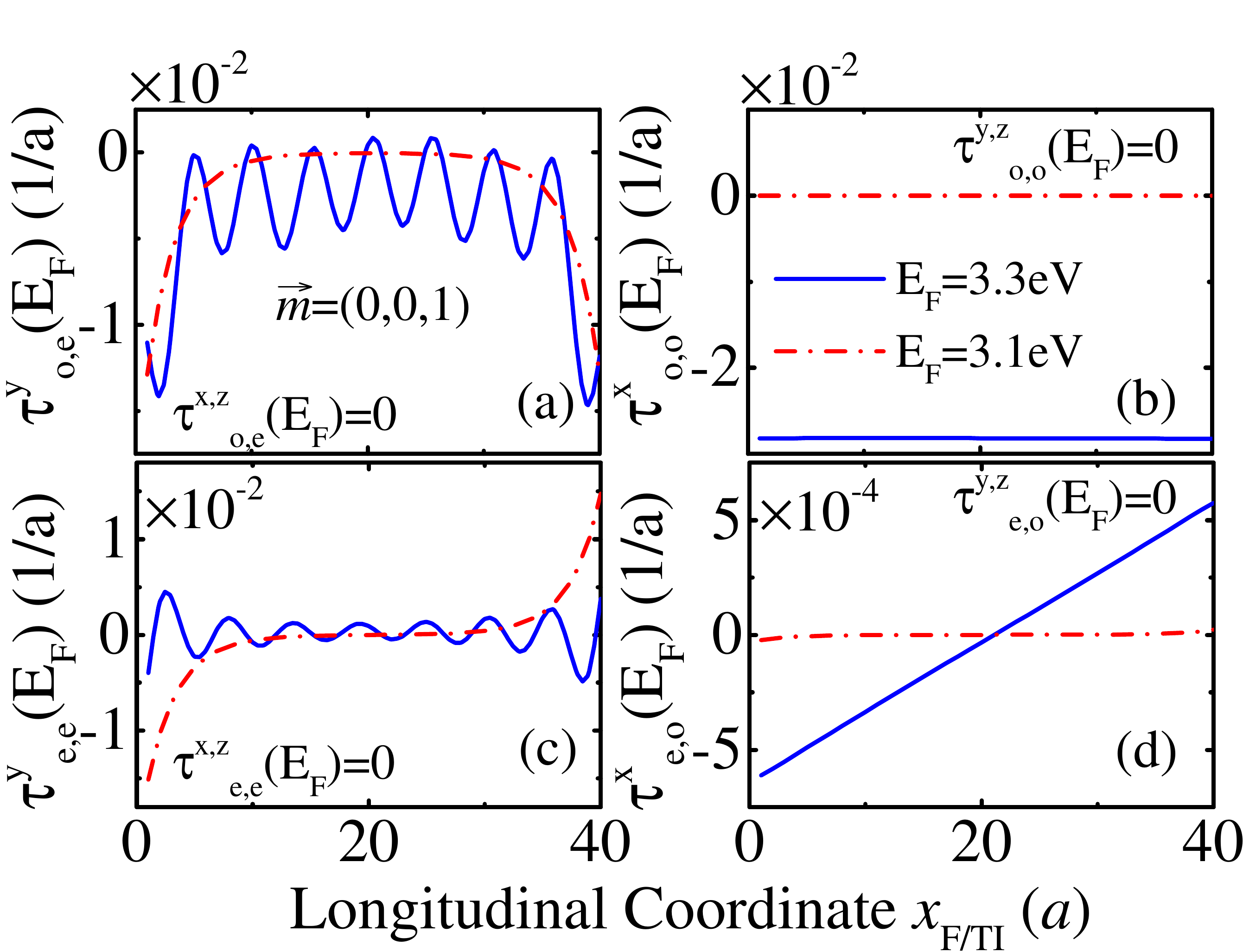}
			\caption{(Color online) (a)--(d) Spatial dependence of SOT components, $\tau^i_{\mu,\nu}(E_F)$ ($i\in\{x,y,z\}$ and $\mu,\nu\in \{e,o\}$), per unit length, for Fermi energy inside the magnetization induced gap around Dirac point ($E_F=3.1eV$) and outside the gap ($E_F=3.3eV$), for \mbox{$\vec{m}\parallel \hat{z}$} in Fig.~\ref{fig:fig1}. Their physical meaning is explained in Fig.~\ref{fig:fig2}. The range of $x$-coordinate corresponds to the length $L_x^F=40a$ of the F overlayer, while the results are independent of the length of the TI layer underneath, $L_x^\mathrm{TI}$.}
			\label{fig:fig4}
\end{figure}

Figure~\ref{fig:fig4}(a) demonstrates that large values of antidamping-like SOT from Fig.~\ref{fig:fig2} are {\it spatially localized} around the transverse edges of the F overlayer, for Fermi energy inside and outside of the surface state gap induced by out-of-plane magnetic exchange coupling. While conductance in this system is close to zero at the Dirac point, we observe that the anti-damping torque does not depend strongly on the Fermi energy. This suggests a high efficiency of SOT per injected current for the Fermi energy close to the Dirac point. For the field-like SOT in Fig.~\ref{fig:fig4}(b) we see the torque independent of the coordinate in the entire F/TI contact region, as expected from the phenomenology of the EE. In Fig.~\ref{fig:fig4}(c) we plot the contribution of the Fermi energy electrons to the FM/TI interface induced MCA field. Even though total $\tau_\mathrm{e,o}^{i}(E_F) \equiv 0$, its spatially-resolved value plotted in Fig.~\ref{fig:fig4}(d) is nonzero which can be removed by performing a proper gauge transformation.  

To understand the origin of the anti-damping SOT let us find an expression for the average of SOT around a fixed axes $\vec{m}=\vec{m}_0+(\vec{m}_{\perp}\cos(\phi)+\vec{m}_0\times\vec{m}_{\perp}\sin(\phi))\delta\theta$ with small cone angle $\delta\theta$. By applying a rotation operator we can align the fixed axes along $z$-axis such that $\vec{m}_0=\hat{e}_z$ and $\vec{m}_{\perp}=\hat{e}_x$. In this case we have
\begin{eqnarray}\label{eq:sot_pert}
\langle{\mathcal{T}}^z\rangle_\phi=\frac{\Delta_{surf}\delta\theta}{2}\Im\int \frac{d\phi}{2\pi}\int \frac{dE}{2\pi}\sum_{\alpha}f_{\alpha}e^{i\phi}T^{-,\alpha},
\end{eqnarray}
where, $T^{-,\alpha}=T^{x\alpha}-iT^{y\alpha}$ and $\Im$ refers to the imaginary part. Expanding the GFs in Eq.~\ref{eq:trans}(b) to the lowest order with respect to $\delta\theta e^{-i\phi}$, we obtain,
\begin{align}
T^{ -,\alpha} = \frac{\Delta_\mathrm{surf}}{4}e^{-i\phi}\delta\theta\Tr&\left[ {\bm 1}_m {\bm \sigma}^- \bold{G}_t{\bm 1}_m {\bm \sigma}^+\bold{G}_t {\bm \Gamma}_{\alpha} \bold{G}_t^\dagger \right.\\
&+\left. {\bm 1}_m {\bm \sigma}^- \bold{G}_t {\bm \Gamma}_{\alpha} \bold{G}_t^\dagger{\bm 1}_m {\bm \sigma}^+\bold{G}_t^\dagger \right].\nonumber
\end{align}
Plugging this expression into Eq.\eqref{eq:sot_pert}, and using the identity, $\bold{G}_t- \bold{G}^\dagger_t=i\sum_{\alpha}\bold{G}_t {\bm \Gamma}_{\alpha} \bold{G}_t^\dagger$, in linear bias voltage regime we obtain,
\begin{eqnarray}\label{eq:sot_pert1}
\langle{\mathcal{T}}^z\rangle_\phi = \frac{V_b \Delta^2_\mathrm{surf} \delta \theta^2}{16} \Tr[{\bm \rho}^{\uparrow\uparrow}_{L}{\bm 1}_m{\bm \rho}^{\downarrow\downarrow}_{R}{\bm 1}_m-{\bm \rho}^{\uparrow\uparrow}_{R}{\bm 1}_m{\bm \rho}^{\downarrow\downarrow}_{L}{\bm 1}_m],\ \ \ \ \
\end{eqnarray}
where, ${\bm \rho}_{\alpha}=-i\bold{G}_{\alpha}^<=\bold{G}_t {\bm \Gamma}_{\alpha} \bold{G}_t^\dagger$, corresponds to the density matrix inside the F overlayer  for the electrons (holes) at the Fermi surface being injected from the lead $\alpha$ ($\beta\neq\alpha$). The electron-hole analogy can be understood by defining the hole density matrix, $i\bold{G}_{\alpha}^>$, from the identity \mbox{$-i(\bold{G}^<_{\alpha}-\bold{G}^>_{\alpha})=2\Im(\bold{G})={\bm \rho}_{\alpha}+\sum_{\beta\neq\alpha}{\bm \rho}_{\beta}$}. By considering left-lead induced holes instead of right-lead induced electrons, we can interpret  Eq.\eqref{eq:sot_pert1} as spin-resolved electron-hole recombination rate, where opposite spins have opposite contributions to the antidamping-like SOT. This picture focuses on the energy anti-dissipative aspect of the phenomena and, since ${\bm \rho}^{\sigma\sigma}_{L}$(${\bm \rho}^{\sigma\sigma}_{R}$) corresponds to the spin-$\sigma$ right (left) moving electrons, Eq.~\eqref{eq:sot_pert1} suggests that spin-momentum locking naturally has a significant effect on the enhancement of the antidamping-like SOT magnitude. In particular, in the case of F/TI interface, the  enhancement of the antidamping-like SOT occurs when the spin-up/down is along the $y$-axis ($\vec{m}_0\parallel \hat{y}$) which is the spin-polarization direction of electrons passing through the surface of the TI induced by the EE. Additionally, in this case the antidamping-like SOT gets smaller away from the F/TI transverse edge because the contribution of both of the leads to the spin density become identical. Therefore the anti-damping torque in this case is more localized around the edge. This effect is more significant when the magnetization is out of the plane and the Fermi energy is inside the $\Delta_\mathrm{surf}\cos \theta$ gap on the TI surface. 
%Eq.\eqref{eq:sot_pert1} also suggests that the anti-damping torque increases when either spin-majority or spin-minority electrons experience Van Hove singularity with strong electron-hole asymmetry. This could explain the sudden increase of the anti-damping torque in Fig.\ref{fig:fig2}(b) from in-plane ($\theta=0^o$) to out of plane magnetization ($\theta=45^o$) when the Fermi energy enters into the surface state gap.

A alternative interpretation of the results can be achieved by considering $\bold{G}_t- \bold{G}^\dagger_t=i\sum_{\alpha}\bold{G}_t^\dagger {\bm \Gamma}_{\alpha} \bold{G}_t$. In this case, the average of the antidamping-like SOT is expressed by
\begin{eqnarray}\label{eq:sot_pert2}
\langle{ \mathcal{T}}^z \rangle_\phi = \frac{V_b}{4}\Tr[{\bm T}^{\uparrow\downarrow}_{LR}-{\bm T}^{\uparrow\downarrow}_{RL}],
\end{eqnarray}
where the F overlayer induced spin-flip transmission matrix is defined as
\begin{eqnarray}\label{eq:sf_trans_mat}
{\bm T}^{\uparrow\downarrow}_{\alpha\beta}=({\bm t}^{\uparrow\downarrow}_{\alpha\beta})^{\dagger}{\bm t}^{\uparrow\downarrow}_{\alpha\beta},
\end{eqnarray}
and
\begin{eqnarray}\label{eq:sf_trans_mat1}
{\bm t}^{\uparrow\downarrow}_{\alpha\beta}=\frac{\Delta_\mathrm{surf}\delta\theta}{2}\sqrt{{\bm\Gamma}_{\alpha}}{\bm G}_t\sigma^+{\bm G}_t\sqrt{{\bm\Gamma}_{\beta}}.
\end{eqnarray}

\begin{figure}
	\includegraphics[scale=0.3,angle=0]{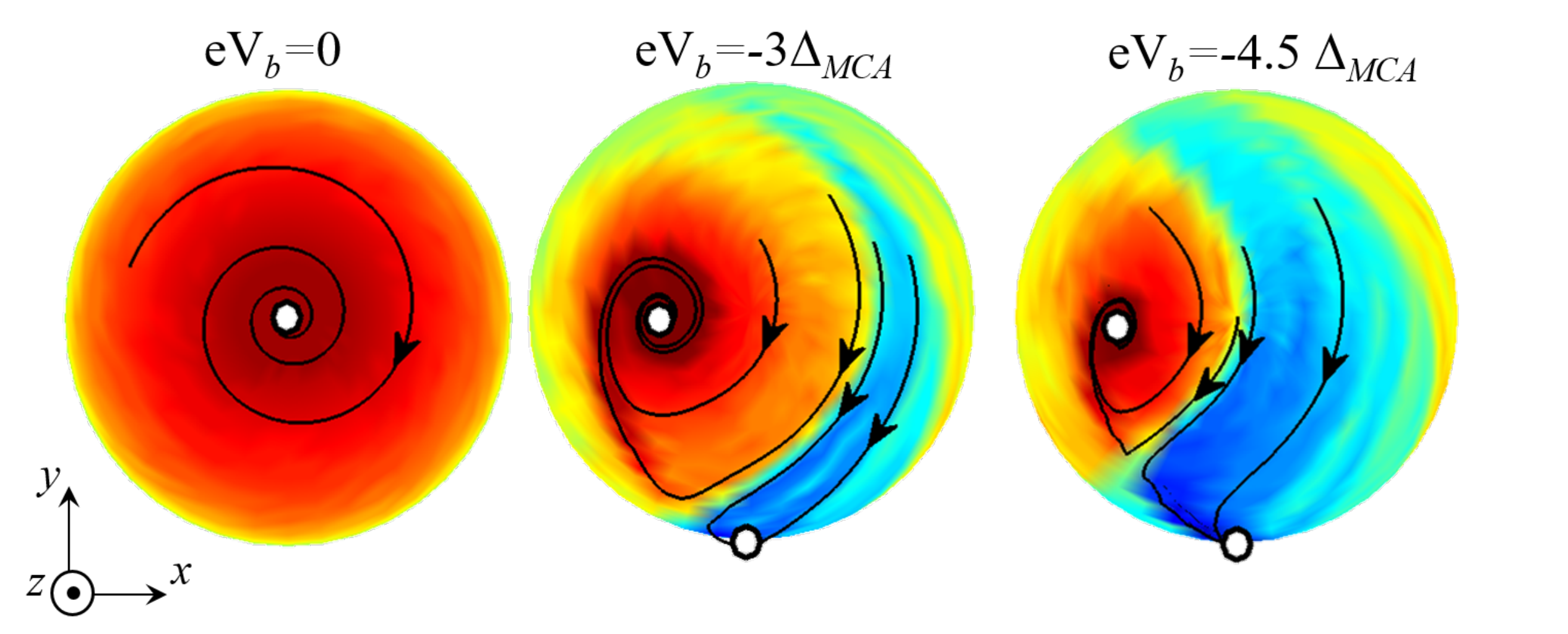}
	\caption{(Color online) SOT-induced magnetization trajectories $\vec{m}(t)$ under different $V_b$ and $\vec{B}^\mathrm{ext}=0$. Higher color intensity denotes denser bundle of trajectories which start from all possible initial conditions  $\vec{m}(t=0)$ on the unit sphere. Solid curves show examples of magnetization trajectories, while the white circles denote attractors of trajectories.}
	\label{fig:fig5}
\end{figure}

Although Eq.~\eqref{eq:sot_pert2} is obtained from perturbative considerations, it looks identical to the Eq.~(8) of Ref.~\onlinecite{Mahfouzi2010} where a spin-flip reflection mechanism at the edge of the F/2D-TI interface was recognized to be responsible for the giant charge pumping (i.e., anti-damping torque) observed in the numerical simulation.~\cite{Mahfouzi2010} Eq.~\eqref{eq:sf_trans_mat1} describes a transmission event in which electrons injected from lead $\alpha$, get spin-flipped (from up to down) by the FM and then transmit to the lead $\beta$. The path of the electrons describing this process is shown in Fig.~\ref{fig:fig1}. From the k-resolved results of the anti-damping torque (not shown here) we observe that while for the in-plane magnetization electrons moving in the same transverse direction (same sign for $k_y$) on both left and right edges of the FM/TI interface contribute to the torque, in the case of out-of-plane magnetization for the left (right) edge of the interface the local anti-damping torque is induced mostly by the electrons with $k_y>0$ ($k_y<0$).

It is worth mentioning that due to nonperturbative nature of the SOT induced by the chiral electrons, the approximation presented in this section which can as well be obtained from the self energy corresponding to the vacuum polarization Feynman diagrams of the electron-magnon coupled system~\cite{Mahfouzi_EM_2014}, does not capture the phenomena accurately. This is evident in the angular dependence of the anti-damping torque which in the current section is considered up to second order effect ($\delta\theta^2$), while the divergence-like behavior in Figs.\ref{fig:fig2}(h) suggest a linear dependence when the magnetization direction is close to the $y$-axis. This signifies the importance of the higher order terms with respect to $\delta\theta$ that can not be ignored. The approximation presented in this section also suggests that blocking the lower surface leads to the reduction of the anti-damping torque. However, in this case an electron experiences multiple spin-flip reflections before transmitting to the next lead and in fact it turns out that the exact results stay intact even if the lower surface is blocked. This is similar to the conclusion made in Ref.~\onlinecite{Mahfouzi2010} which shows the redundancy of blocking the lower edge of the 2D-TI to obtain a nonzero pumped charge current from precessing FM as proposed in Ref.~\onlinecite{Qi2008}.

Although spin-momentum locking of the surface state of the TI resembles the 2D Rashba plane, in the case of TI surface state the cones with opposite spin-momentum locking reside on opposite surface sides of the TI slab while in the case of a Rashba plane they are only separated by the SOC energy. This means one can expect a smaller SOT for a FM on top of a 2D Rashba plane due to cancellation of the effects of the two circles with opposite spin-momentum locking, where the nonzero anti-damping torque originates from the electron-hole asymmetry.

%The introduction of the spin-flip reflection mechanism, that happens when electrons enter into the FM/TI interface, to interpret the results signifies the necessity of electron scattering from the FM moments as the source of the anti-damping torque. This shows that to have a significant anti-damping torque in the middle of the interface we need to introduce impurity of interfacial roughness. To check this conjecture we calculate the anti-damping torque versus the roughness of the interface and present the result in Fig.~\ref{fig:fig7}. In this calculation the roughness was introduced through addition of a large onsite potential to random positions at the interface and then the results are averaged over different configurations of the onsite potential.

%\begin{figure} 
%	\includegraphics[scale=0.25,angle=0]{fig7}
%	\caption{(Color online) .}
%	\label{fig:fig7}
%\end{figure}

\section{LLG simulations of magnetization dynamics in the presence of SOT}\label{sec:llg}

In order to investigate ability of predicted antidamping-like SOT to switch the magnetization direction of a perpendicularly magnetized F overlayer in the geometry of Fig.~\ref{fig:fig1}, we study magnetization dynamics in the macrospin approximation by numerically solving LLG equation at zero temperature supplemented by SOT components analyzed in Sec.~\ref{sec:angular}
	\begin{align}\label{eq:LLG2}
	 \frac{\partial \vec{m}}{\partial t}= & \frac{1}{2 \pi} [\vec{\tau}_\mathrm{o,e}(\vec{m},E_F)+\vec{\tau}_\mathrm{o,o}(\vec{m},E_F)] eV_{b} +\gamma\vec{B}^{ext}\times \vec{m} \nonumber \\
	+& \vec{m} \times \left[ \boldsymbol{\alpha}(\vec{m})\cdot\frac{\partial \vec{m}}{\partial t} \right] + (\vec{m}\cdot\hat{z})(\vec{m}\times \hat{z})\Delta_\mathrm{MCA}.
	\end{align}
Here $\gamma$ is the gyromagnetic ratio, $\boldsymbol{\alpha}(\vec{m})_{ij}= \Delta_\mathrm{surf}^2 T^{ij}(\vec{m},E_F)/8\pi$ is the dimensionless Gilbert damping tensor, and \mbox{$\Delta_\mathrm{MCA} =\Delta^0_\mathrm{MCA}+|\vec{\mathcal{T}}_\mathrm{e,e}|/|(\vec{m}\cdot\hat{z})(\vec{m}\times \hat{z})|$}, where $\Delta^0_\mathrm{MCA}$ represents the intrinsic MCA energy of the FM. We solve Eq.~\eqref{eq:LLG2} by assuming that the Gilbert damping is a constant (its dependence on $\vec{m}$ is relegated to future studies) and ignore the dependence of $\Delta_\mathrm{MCA}$ on $\vec{m}$ and $V_b$ while retaining its out-of-plane direction. 

Figure~\ref{fig:fig5} shows the magnetization trajectories for all possible initial conditions  $\vec{m}(t=0)$ on the unit sphere under different $V_b$. At $V_b=0$, the two attractors are  located as the north and south poles of the sphere. At finite $V_b$, the attractors shift away from the poles along the $z$-axis within the $xz$-plane, while additional attractor appears on the positive (negative) $y$-axis under negative (positive) $V_b$. Note that the applied bias voltage $V_b$ drives dc current and SOT proportional to it in the assumed linear-response transport regime.

\begin{figure}
		\includegraphics[scale=0.41,angle=0]{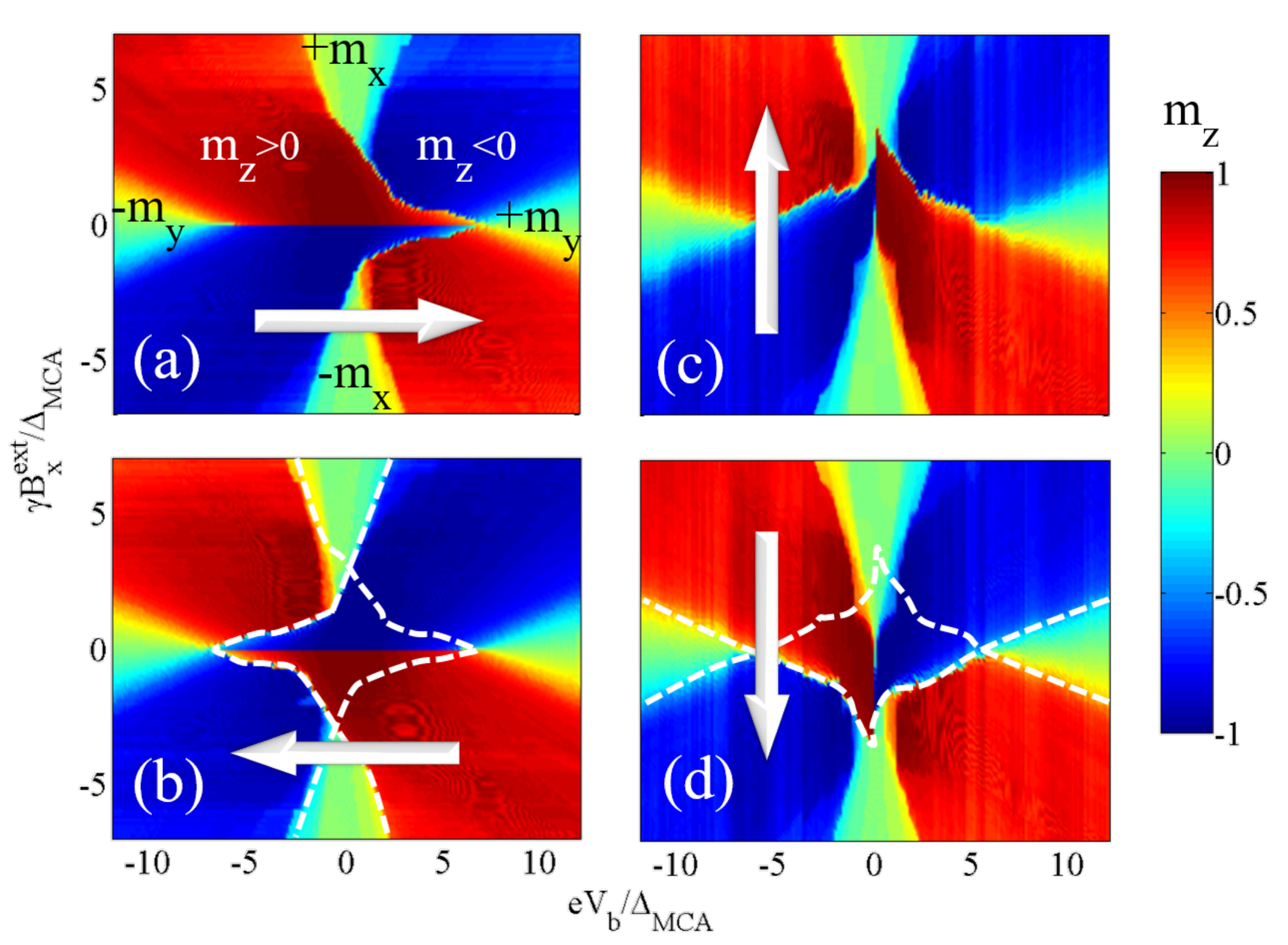}
		\caption{(Color online) Phase diagram of the magnetization state in lateral F/TI heterostructure from Fig.~\ref{fig:fig1} as a function of an in-plane external magnetic $\mathbf{B}^\mathrm{ext} \parallel \hat{x}$ and $V_b$ (i.e., SOT $\propto V_b$). Thick arrows on each of the panels (a)--(d) show the direction of sweeping of  $B^\mathrm{ext}_x$ or $V_b$ parameter. The smallness of central hysteretic region along the $V_b$-axis, enclosed by white dashed line in panel (b) and (d), shows that low currents are required to switch  magnetization from $m_z>0$ to $m_z<0$ stable states.}
		\label{fig:fig6}
\end{figure}

Figure~\ref{fig:fig6} shows the commonly constructed~\cite{Fan2014a,Liu2012,Liu2012c,Ndiaye2015} phase diagram of the magnetization state in the presence of an  external in-plane magnetic field $\mathbf{B}^\mathrm{ext} \parallel \hat{x}$ and the applied bias voltage $V_b$ (i.e., SOT $\propto V_b$). The thick arrows in each panel of Fig.~\ref{fig:fig6} denote the direction of the sweeping variable---in Fig.~\ref{fig:fig6}(a) [\ref{fig:fig6}(b)] we increase [decrease] $V_b$ slowly in time, and similarly in Fig.~\ref{fig:fig6}(c) [\ref{fig:fig6}(d)] we increase [decrease] the external magnetic field gradually. The size of hysteretic region in the center of these diagrams, enclosed by white dashed line in Figs.~\ref{fig:fig6}(b) and ~\ref{fig:fig6}(d), measures the efficiency of switching.~\cite{Fan2014a,Liu2012,Liu2012c,Ndiaye2015} Since this region, where both magnetization states $m_z>0$ and $m_z<0$ are allowed, is relatively small in Figs.~\ref{fig:fig6}(a) and ~\ref{fig:fig6}(b), magnetization can be switched by low $B_x^\mathrm{ext}$ and small $V_b$ (or, equivalently, small injected dc current), akin to the phase diagrams observed in recent experiments.~\cite{Fan2014a} 

Although we considered the FM a single domain, the fact that the anti-damping component of the SOT is mainly peaked around the edge of the FM/TI interface suggests that it is be more feasible in realistic cases to have the local magnetic moments at the edge of the FM switch first and then the total magnetization switches by the propagation of the domain walls formed at the edge throughout the FM~\cite{Yu2014,Mikuszeit2015}. Therefore, a micromagnetic simulation of the system is required to investigate switching phenomena in large size systems which we relegate to future works. 

\section{Conclusions}\label{sec:conclusion}

	In conclusion, by performing adiabatic expansion of time-dependent NEGFs,~\cite{Stefanucci2013,Haug2008} we have developed a framework which yields formulas for spin torque and charge pumping as reciprocal effects to each other connected by time-reversal, as well as Gilbert damping due to SOC. It also introduces a novel way to separate the SOT components, based on their behavior (even or odd) under time and bias voltage reversal, and can be applied to arbitrary systems dealing with classical degrees of freedom coupled to electrons out of equilibrium. For the geometry~\cite{Sanchez2013} proposed in Fig.~\ref{fig:fig1}, where the F overlayer  covers (either partially or fully) the top surface of the TI layer, we predict that low charge current flowing solely on the surface of TI will induce antidamping-like SOT on the F overlayer via a physical mechanism that requires spin-momentum locking on the surface of TIs---spin-flip reflection at the lateral edges of a ferromagnetic island introduced by magnetic proximity effect onto the TI surface. This mechanism has been overlooked in efforts to understand why SO-coupled interface alone (i.e., in the absence of SHE current from the bulk of SO-coupled non-ferromagnetic materials) can generate antidamping-like SOT, where other explored mechanisms have included  spin-dependent impurity scattering at the interface,~\cite{Pesin2012a} Berry curvature mechanism,~\cite{Lee2015,Li2015} as well as their combination.~\cite{Qaiumzadeh2015} 
	
	%The spin-flip reflection mechanism in the geometry of Fig.~\ref{fig:fig1}  finesses a major drawback of F/TI heterostructures explored in recent experiments~\cite{Mellnik2014,Fan2014a,Wang2015} where most of the applied current was shunted through the metallic F overlayer and, therefore, did not contribute to SOT. We emphasize that F overlayer in Fig.~\ref{fig:fig1} can be either metallic or 	insulating, as long as it can introduce the proximity exchange field in Eq.~\eqref{eq:proximity} onto the TI surface. This is confirmed by observing that our predictions for SOT are virtually unchanged as we increase the number of monolayers $L_z^F$ of the metallic F layer attached to the surface of TI. 
	The key feature for connecting experimentally observed SOT and  other related phenomena in F/TI heterostructures (such as spin-to-charge conversion~\cite{Sanchez2013,Mahfouzi2014a,Shiomi2014}) to theoretical predictions is their dependence~\cite{Garello2013,Jamali2015} on the magnetization direction. The antidamping-like SOT predicted in our study exhibits complex angular dependence, exhibiting ``nonperturbative'' change with the magnetization direction in Fig.~\ref{fig:fig2}(h), which should make it possible to easily differentiate it from other competing physical mechanisms.

\begin{acknowledgments}
 F. M. and N. K. were supported by NSF PREM Grant No. 1205734, and B. K. N. was supported by NSF Grant No. ECCS 1509094.
\end{acknowledgments}

 %BibTeX
 	%Windows:
 	%\bibliography{D:/PHYSICS/TEX/BIBTEX/qttg}
   %\bibliographystyle{D:/PHYSICS/TEX/BIBTEX/utphys.bst}

\end{document}